\documentstyle[twocolumn,aps,epsf]{revtex}
\begin{document}
\title{A thermostable trilayer resist for niobium lift-off} 
\author{P. Dubos$^{1}$, P.  Charlat$^{1}$, Th.  Crozes$^{1}$, P. Paniez$^{2}$ and 
B. Pannetier$^{1}$} 
\address{$^{1}$Centre de Recherches sur les Tr\`es 
Basses Temp\'eratures, C.N.R.S.,\\ Laboratoire conventionn\'e avec 
l'Universit\'e Joseph Fourier, \\
F-38042 Grenoble, France.\\
$^{2}$France Telecom, C.N.E.T.- Grenoble, DTM/TFM, P.O. Box 98 \\
F-38243 Meylan Cedex, France}

\maketitle

\begin{abstract}
We have developped a novel lift-off process for fabrication of high 
quality superconducting submicron niobium structures. The 
process makes use of a thermostable polymer with a high transition 
temperature
$T_{g}= $ 235$^\circ$C and an excellent chemical stability. The superconducting 
critical temperature of 100 nm wide niobium lines is above 7 K. An example of 
shadow evaporation of a Nb-Cu submicron hybrid structure is given.  A 
potential application of this process is the fabrication of very small 
single electron devices using refratory metals.
\end{abstract}
PACS index category: 74.80 Fp, 85.25, 85.40

\section{Introduction}

The field of single charge tunneling phenomena \cite{houches}, mesoscopic 
superconductivity \cite{Meso} or superconducting devices 
\cite{bolometers} has opened a new demand in high performance 
nanofabrication techniques with superior self-alignement capabilities. 
A common technique makes use of shadow 
evaporation through a suspended stencil mask prepared by electron beam 
lithography\cite{Dolan}. This technique allows self-alignement with nanometer 
scale accuracy as required for fabrication of ultrasmall tunnel 
junctions.  Excellent results are currently obtained using 
conventional techniques based upon masks with PolyMethylMethAcrylate 
(PMMA) as the e-beam sensitive resist.  The high resolution stencil 
mask is formed on top of a sub-layer with a well controlled undercut. 
The mask is easy to remove by a lift-off process.  In the 
conventional two-layers process \cite{Beaumont,Howard} the upper layer 
is a thin layer of PMMA (casting solvent Chlorobenzene) while the 
bottom layer is a copolymer PMMA-MAA containing MethAcrylic Acid (MAA) 
monomers.  These MAA co-monomers make the copolymer soluble in polar 
solvents such as acetic acid and provide good chemical selectivity 
with respect to the top layer.  Alternately the stencil mask can be 
made of a thin film of germanium or silicon patterned using an 
additionnal PMMA top layer (tri-layers process).  Such a process, as 
well as more complex alternative processes with more than three 
layers, generally ensures excellent control of each intermediate step.  
It is widely used for the fabrication of devices made of soft 
materials such as aluminum, gold, copper, chromium, permalloy, etc.  
Structures of high complexity can be realized by multiple angle 
evaporation using one or two rotation axis\cite{Courtois}.

Unfortunately, this technique cannot be extended to refractory 
metals such as niobium, molybdenum, tungsten or tantalum which require 
both high vacuum and high evaporation temperatures. As a consequence 
of the excessive heat produced by the electron beam evaporation, conventional 
resist masks are mechanically unstable. In addition, contamination 
due to the resulting outgasing of the resist degrades the electronic 
properties of the metal. It is a well known fact that the 
superconducting properties of niobium are extremely 
sensitive to a small amount of oxygen contamination\cite{oxygen}.  
Various methods have been attempted to extend the shadow evaporation 
technique to refractory metals with more stable mask structures.  In 
Ref.  \cite{vdz91}, a combination of chromium mask with a metallic 
aluminum sub-layer was successfully used to fabricate arrays of micron 
size niobium/niobium Josephson junctions.  More recently, Harada et 
al\cite{Harada} developped a four layer resist system composed of 
PMMA, a hard baked photoresist, germanium and PMMA.  This process was 
used successfully to fabricate submicron niobium/aluminum 
oxyde/niobium superconducting single electron transistors.  However, 
the measured critical temperature of the niobium electrode was far 
below that of the bulk material ($9.2 K$) and therefore the device 
could not reach optimum operation.\\
In this paper, we describe a new process based upon a thermostable 
polymer, the Poly PhenyleneEtherSulfone (PES) which greatly improves 
the quality of the devices.  We present a detailed 
comparison between the thermal characteristics of this polysulfone and that 
of PMMA polymer.  We show in particular why the latter should be avoided as a 
sub-layer resist.  As a demonstrator for this process we have 
fabricated submicron niobium/copper/niobium Josephson junctions. 

\section{Selection of the thermostable base layer for the trilayer process}

We have chosen to develop a new trilayer process with a thermostable 
polymer as the base layer and silicon (alternately germanium) as the 
high resolution stencil mask.  The thermal stability of the PMMA upper 
layer is irrelevant since this layer only serves for patterning the 
silicon mask and is removed before the evaporation process.  In order 
to select an alternative to PMMA (or PMMA-MAA) as the bottom layer, we 
have explored a number of thermostable polymers of the phenolic 
family.

The first series of experiments was carried out on a PHS polymer (Poly 
ParaHydroxyStyrene) whose chemical formula is $(CH_{2}-CHX-)_{n}$ 
where X is the phenolic group.  This polymer is a polystyrene with an 
hydroxyl function which insures solubility in polar solvents.  Three 
differents molecular weight have been used: 23.600, 30.000 and 109.000 
g/mole.  The powder was diluted in "Diglyme" (2-Methoxyethyl Ether).  
The glass transition temperature of this polymer is about 
180$^\circ$C.  It can be safely used at temperatures below 
240$^\circ$C.  Above this temperature, cross-linking makes the polymer 
insoluble.  Good results were obtained with this polymer as a bottom 
layer in the trilayer process.  However, it turned out to be damaged by the 
solvents used for the development and rinsing of the PMMA upper layer.

\begin{figure}
\epsfxsize=8 cm \epsfbox{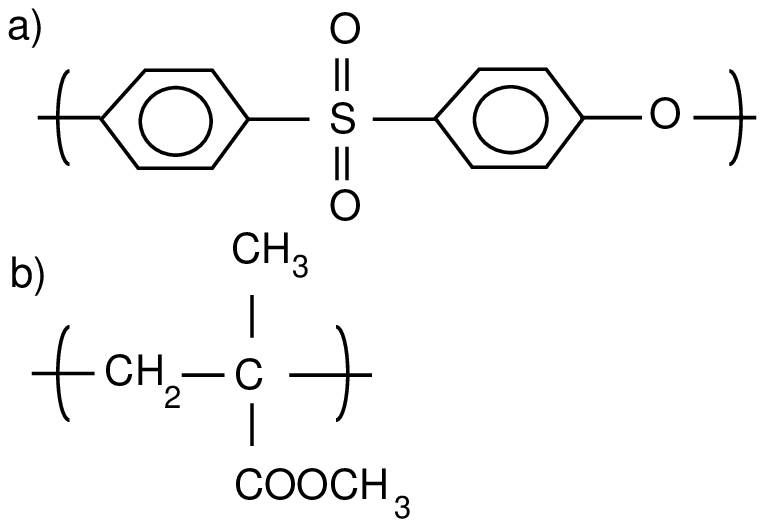}
\caption{a) The PES is a sequence of aromatic groups attached 
to a Sulfur atom. b) The PMMA chemical formula : the weak bond between 
monomers explains its high sensitivity to e-beam lithography and temperature.}
\label{formule}
\end{figure}

The best results were actually obtained with a polymer presenting a lower 
solubility: the poly PhenyleneEtherSulfone (PES) originally commercially 
available under the name Victrex from ICI.  This polymer is currently used 
as a thermostable organic substance for industrial use at 
180$^\circ$C.  The chemical formula of PES is a sequence of aromatic 
groups attached to a sulfur atom.  The monomer has the structure shown 
in Fig. \ref{formule}a.

The high thermal stability of PES is insured by the aromatic groups. 
We have used a Poly PhenyleneEtherSulfone with a high molecular weight: PES 
Victrex 5003P (ICI).  The PES is dissolved in N-methyl Pyrrolidone 
($10\%$ w/w) to form a solution that yields a
standard thickness of a few hundred nanometers after spinning. For comparison 
the monomer of PMMA is described in Fig. \ref{formule}b. The 
chemical bond between MMA monomers is weak. This is the reason for the 
excellent properties of PMMA as a high resolution and good sensitivity 
electron beam resist as well as its moderate thermal stability.  \\
The chemical properties of PES are also very attractive since 
this polymer exhibits a good compatibility with the organic solvents 
used in the subsequent steps of the trilayer process.  It is inert in 
the PMMA solvents - Chlorobenzene or Methyl-Iso-Butyl Ketone (MIBK) 
and is insoluble in IsoPropylic Alcohol.  Its most convenient solvents 
are the DiMethyl Sulfoxyde (DMSO) and the N-Methyl Pyrrolidone (NMP).  
The latter is much less dangerous and volatile than 
Chlorobenzene which is a standard solvent of PMMA.

\section{Thermal characteristics of the base resist layer}

The thermal properties of PMMA, PMMA-MAA and PES polymers were 
investigated using a TA Instruments 2950 Thermogravimetry Analyser (TGA) 
and a TA Instruments 2920 Differential Scanning Calorimetry (DSC) 
Analyser. The heating ramp for experiments was 10$^\circ$C/min.

\begin{figure}
\epsfxsize=8 cm \epsfbox{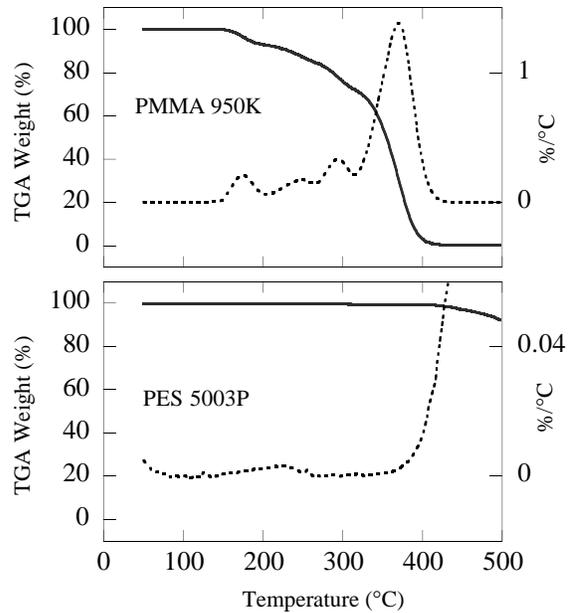}
\caption{Upper graph: TGA results for PMMA polymer with a molecular weight 
$M_{W}= 950.000g/mole$.  Lower graph: TGA 
results for PES polysulfone. The full line indicates the weight loss vs. 
temperature in percent.  The dashed lines represent the derivative of 
the TGA curves and indicate degassing of volatile species which are particularly 
abundant in PMMA.}
\label{tga2_pes_pmma}
\end{figure}

The enhanced thermostability of PES 
is illustrated in Fig. \ref{tga2_pes_pmma} (lower graph) which 
demonstrates that its weight loss is negligible at temperatures 
up to 400$^\circ$C, while a significant loss of weight is observed in 
PMMA at only 150$^\circ$C.  This temperature can easily be reached 
during e-beam evaporation of a refractory metal.  The peaks in the 
derivative curve (dashed line) indicate the activation temperatures 
for chemical transformations such as anhydride formation or 
hydrocarbon outgassing which are important at moderate temperatures 
for PMMA. This polymer presents an extremely high chemical 
stability between 275$^\circ$C (hard-bake temperature) and 
400$^\circ$C.  Above this temperature, the polymer properties degrade 
sharply.

\begin{figure}
\epsfxsize=8 cm \epsfbox{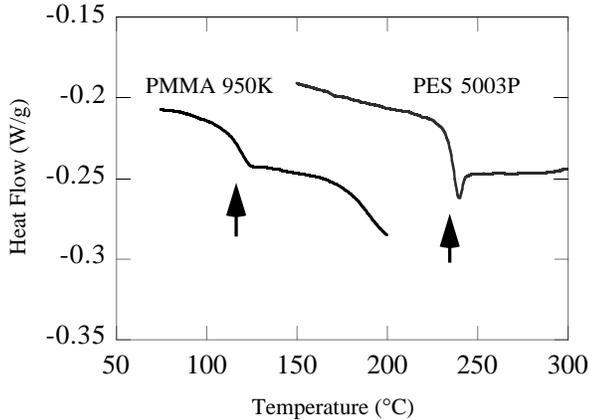}
\caption{Comparison of Differential Scanning Calorimetry (DSC) results 
for PES 5003P and PMMA 950K polymer.  The arrows indicate the glass 
transition temperatures.}
\label{dsc}
\end{figure}

Fig.  \ref{dsc} shows a Differential Scanning Calorimetry (DSC) 
thermogram of the two polymers.  We found a glass transition 
temperature of respectively 235$^\circ$C for PES and 121$^\circ$C for 
PMMA, again showing the superior thermal properties of PES.

In general PMMA as well as its copolymers \cite{Moreau} are well known 
to exhibit poor thermal properties. In particular the PMMA-MAA copolymer 
exhibits low $T_{G}$ (133$^\circ$C for 8.5 $\%$ w/w of acid) and a
high rate of weight loss at moderate 
temperatures\cite{Paniez}. As a result, a very strong outgassing 
of the resist bottom layer takes place in the vicinity of the device, 
even though no pressure increase was recorded by the vacuum gauge in 
our experiment. Indeed, 
we have observed that the niobium structures evaporated in a ultra-high 
vacuum chamber (base pressure of the chamber $10^{-10}$ mbar, sample 
at 25 cm from niobium heated target) through a PMMA/PMMA-MAA bilayer, were 
not superconducting at $1K$.  We believe that the niobium film traps 
moisture, oxygen and hydrocarbures outgassed from the heated polymer 
sub-layer.  We should mention that, with careful limitation of heat 
radiation eg. using copper shields, sequential evaporation and liquid 
nitrogen cooling of the substrate holder, intermediate superconducting 
critical temperatures could be achieved with a PMMA sub-layer.  
Additional improvement could even be obtained by encapsulating the 
PMMA resist with either silicon or aluminum oxyde.  As discussed 
below, the process based upon the PES thermostable sublayer is free of 
these constraints.

 \section{Process Implementation}

The optimized trilayer fabrication process is as follows: Firstly, 
the PES 5003P solution ($10\%$ w/w in NMP) was spinned on for 5 min 
at 2000 rpm on a 2 inch silicon wafer to form the 300 nm thick bottom layer.
After baking at 275$^\circ$C for one minute on a hot-plate, a 40 nm thick silicon layer was 
e-beam evaporated at room temperature on the PES bottom layer.  Alternately, 
germanium could be used without any change in the rest of the process.  
Finally a 85 nm thick PMMA layer ($2\%$ w/w PMMA 950K in
chlorobenzene) was spun-on on top of the silicon.  This thin PMMA layer 
was then patterned by electron beam lithography using a modified 
Scanning Electron Microscope Cambridge S240 with homemade interfaces 
for e-beam writing.  Subsequently, the PMMA layer was developped in a 
solution (1:3) of MIBK and IsoPropylic Alcohol (IPA) for 20 seconds.  
The unprotected silicon was then removed by a Reactive Ion Etching 
(R.I.E.) in a Plassys MG200 reactor.  The etching parameters were: 15 
sec with 20 $sccm$ $SF_{6}$ at pressure $2.10^{-2} mbar$, incident RF 
power 20 $W$, chamber at 15$^\circ$C.  We found it useful to place the 
substrate on a $4"$ pure silicon wafer in order to increase the 
etching time and optimize its reproducibility.  Depending on the 
desired depth of undercut, two alternative processes could be used to 
etch the PES bottom layer \cite{electrosensitivity}:

\begin{figure}
\epsfxsize=8 cm \epsfbox{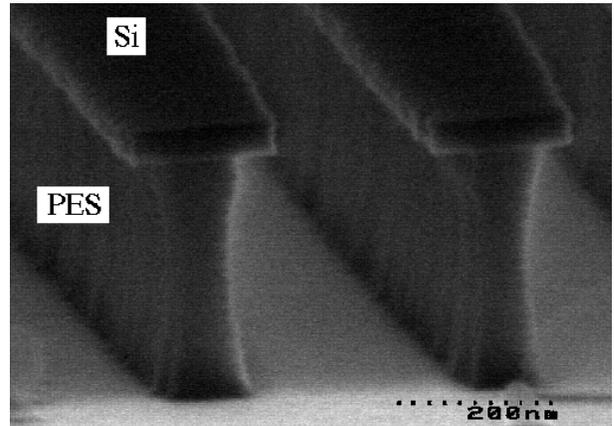}
\caption{Small undercut obtained by a dry process : R.I.E.  oxygen for 
3 min.  The undercut is less than 50 nm.  The walls of sub-layer PES resist are 
320 nm high and 100 nm wide.}
\label{sec}
\end{figure}

\begin{itemize}
\item   small undercut :
To obtain undercuts below 50 nm we used a dry process which consisted 
of an oxygen R.I.E.  in the same reactor as above (but without 
the $4"$ pure silicon wafer).
The etching conditions were the following : 3 min, 20 sccm oxygen at 
pressure $4.10^{-1}$ mbar, incident RF power $50W$ and chamber 
temperature at 15$^\circ$C. An example of this dry process is shown in 
Fig. \ref{sec}.  This process can be extended to produce larger undercuts.
However, residues of typical size 50 nm were usually obtained on the substrate.
These residues are strongly resistant to R.I.E. and could only be eliminated 
using a wet process followed by additional oxygen plasma.

\begin{figure}
\epsfxsize=8 cm \epsfbox{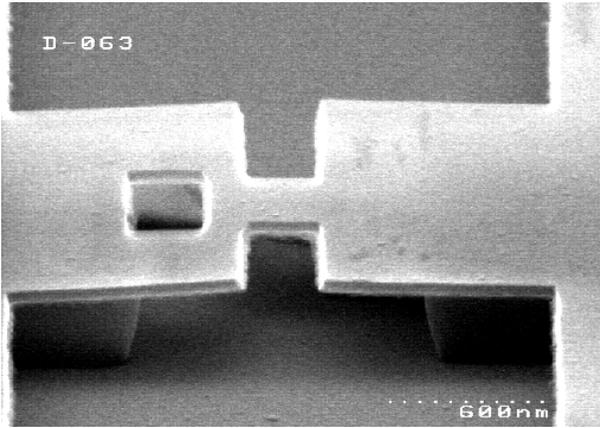}
\caption{Large undercut obtained in DMSO solvent. The PES resist is 
over-etched on 450 nm on both sides of a 300 nm wide hole in Silicon 
mask}

\label{humide}
\end{figure}

\item large undercut : 
To obtain undercuts larger than 50 nm a wet process based on DiMethyl 
Sulfoxide (DMSO) was found to be better.  Thus, excellent control of both 
the etching time and temperature are extremely important since the 
etch rate is thermally activated.  The sample was dipped and shaked 
into DMSO solvent kept at its melting temperature (18.6$^\circ$C).  We 
found that the undercut obeys a quasi-linear dependence as a function 
of etching time (see Table \ref{tableDMSO}) with a typical rate of 28 
$nm/ sec$.  Ethanol was used both to stop the etching and to rinse the 
sample.  An example of this wet process is shown in Fig.\ref{humide}.
\end{itemize}

\begin{table}[tbp]
\begin{tabular}{cccc}
 \hspace{4cm} & \hspace {4cm}\\
        Time & Undercut \\ \hline \hline
        5 sec & 120 nm \\
       10 sec & 280 nm \\
       15 sec & 450 nm \\
       20 sec & 650 nm
\end{tabular}
\caption{Undercut width vs. time for wet process when sample is dipped and 
shaked in pure DMSO solvent.}
\label{tableDMSO}
\end{table}

Let us now discuss some practical points related to PES :

{\it Humidity sensitivity :} The control of ambient humidity is 
crucial during the spinning of the 
polysulfone. The relative humidity level in a standard clean-room 
is about $50\%$. In these conditions, the visual aspect 
of the resist surface after spinning is grey with white points due to 
local phase inhomogeneity. 
The upper limit appears to be $25\%$ of relative humidity during 
the spin coating step. A convenient solution consists of 
drying the atmosphere by blowing a dry nitrogen flow in a small 
chamber surrounding the spinner.  We use a 6 $dm^{3}$ cylindrical 
plexiglass chamber which contains an hygrometer, a pipe for nitrogen 
flow and a small aperture on the top to inject the resist with a poly 
Propylene syringe onto the sample.  The relative humidity is reduced 
to below $15\%$ in a few minutes.  Nitrogen flow
was maintained during spinning. This also accelerates the evaporation of NMP 
solvent from the PES layer.  An homogeneously colored
surface is then easily obtained.

{\it Ageing sensitivity :} Without special storage precaution (humidity-free 
room), trilayers have to be used within 3 months of preparation.  After 
this time, the PES layer could not be undercut in the specific solvent 
given below.

{\it HydroFluoric acid sensitivity :} PES polymer shows an excellent stability 
against HydroFluoric acid (HF) rinsing. It allows preparation of the substrate silicon surface 
through the mask before the metallic evaporation. HF rinsing was used in order 
to remove native silicon oxide. It ensures a
high-quality deposited metal as it decreases 
hydrocarbon contaminants and passivates chemically the silicon surface. The 
major result was an excellent sticking of the thin 
lithographic structures after the final lift-off step. The surface 
preparation consists of 1 minute shaking of the final stencil mask in a 
solution of $10\%$ in volume of HF and a 10 minutes rinsing in 
desionised
water to remove completely the acid.

\section {Test device}

Various niobium devices have been fabricated using the above described 
process. The niobium was evaporated 
at room temperature from an electron beam gun in a ultra high 
vacuum chamber.  No deformation of the silicon mask was observed after 
the electron beam evaporation.  The stencil mask was lifted-off in 
NMP at 80$^\circ$C for 10 mins followed by a few seconds in low power 
ultrasound (NEY ultrasonik 300).  We measured the resistance of all 
these niobium structures using four-probes measurement with standard 
lock-in techniques at low temperature.\\
 
To validate the recipe, thin niobium wires were realised.  We first 
designed the mask implementing a dry etching step in the process as 
described above.  The geometry was a single metallic line of 5 $\mu$m 
long, 0.3 $\mu$m wide and 60 nm thick.  The critical superconducting 
temperature obtained was 7.2 K (see first line of Table \ref{samples}).  
Another mask using wet etching step for fabrication was tested.  The 
structure was an array (size 10 $\mu$m x 5 $\mu$m, step 1 $\mu$m) of 
0.15 $\mu$m wide lines.  60 nm thick niobium were evaporated through 
this stencil mask.  The critical temperature obtained was 7.1 K with a 
residual resistivity ratio (RRR) of 1.6.  The same array with 0.35 
$\mu$m wide lines exhibit a $T_{c}$ of 8.1 K (see line 2 and 3 in 
Table \ref{samples}).  These results may be compared to a measurement on a 
reference plain niobium film of same thickness.  In such a case, 
$T_{c}$ is about 8K with a RRR of 1.86.

\begin{table}[tbp]
\begin{tabular}{ccccc}
 \hspace{1cm} & \hspace {1cm} & \hspace {1cm} & \hspace {1cm} & \hspace 
 {1cm} \\
        Process & RRR & w ($\mu$m) & e (nm) & $T_{c}$ (K) \\ \hline \hline
        dry &     & 0.3   & 60 & 7.2 \\
        wet & 1.6 & 0.15  & 60 & 7.1 \\
        wet &     & 0.35  & 60 & 8.1 \\
        reference & 1.86  & & 60 & 8
\end{tabular}
\caption{Parameters (dry or wet process, resistivity ratio, 
line width, thickness and critical temperature) of the niobium devices.}
\label{samples}
\end{table}

\begin{figure}
\epsfxsize=8 cm \epsfbox{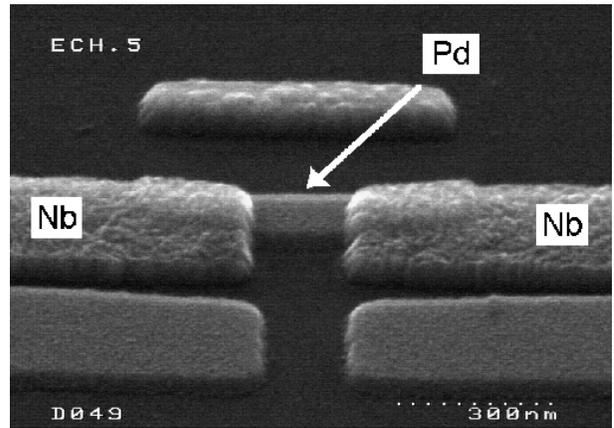} 
\caption{Micrograph (oblique 
view) of a niobium-palladium-niobium SNS junction realised by shadow
evaporation with high transparency S-N interfaces.}
\label{SNS}
\end{figure}

A geometry currently studied in mesoscopic physics is the 
superconducting - normal metal - superconducting SNS structure where 
the N island is viewed as a "quantum dot" coupled to superconducting 
electrodes through the Andreev process \cite{Beenakker}. We are studying 
such SNS junctions with highly transparent interfaces\cite{dubos}. The 
mask shown in Fig. \ref{humide} allows the in-situ fabrication of 
self-aligned SNS junctions by shadow evaporation using copper or 
palladium as the normal metal island and niobium as the superconductor. 
The alignement of the normal island is achieved  with a nanometer 
scale resolution by the proper choice of the evaporation angle. High quality S-N 
interfaces free of contamination was ensured as the metals were evaporated 
within one cycle in a ultra high vacuum (UHV) chamber. A typical sample
with palladium as the normal metal is shown in Fig. \ref{SNS}.  We are 
making systematic measurements at low temperature
of niobium-copper-niobium SNS junctions have been done. Interface resistance 
were estimated below 0.2 $\Omega$.  Niobium lines were 0.3 $\mu$m wide 
and 60 nm thick and their resistivity was in the range of 17-20 
$\mu\Omega$.cm.  Copper metal was 0.3 $\mu$m wide and 60 nm thick for a 
resistivity of 4.3 $\mu\Omega$.cm.  In such a SNS junction the 
superconducting critical 
temperature was above 7K.  Because the alignement was made in-situ under 
high vacuum, this technique also allows an excellent control of the 
interface between the central island and the external superconducting 
electrodes : from the metallic contact (high transparency barrier) to 
the weak tunnel junction (low transparency barrier).  Masks with large 
undercuts (ses Fig. \ref{humide}) can also serve to elaborate niobium based tunnel 
nano-junctions.  Controlled oxydation of artificial or natural tunnel barriers can be 
performed in the UHV chamber between the two evaporations ensuring both 
high quality barriers and a high energy gap for niobium.

\section{Conclusion}

We have demonstrated a reliable technique to produce high resolution 
self-aligned structures by shadow evaporation of refractory metals. 
The key point is the use of a trilayer process with a thermostable 
resist bottom layer. The above recipe has been successfully tested on submicron 
niobium copper mesoscopic structures with excellent superconducting 
properties of the niobium film and excellent control of the interfaces. 
Using this technique we have also
fabricated niobium microsquid gradiometers.  This process is very 
promising in the area of single electronic transistors since it makes 
the shadow evaporation technique accessible to new materials with 
superior electronic properties.

\section{Acknowledgements}
We would like to acknowledge discussions with Th. Fournier, O. Buisson and 
H. Courtois.  We would like to thank D. Mariolle and F. Martin for 
the S.E.M.  micrographs taken at LETI-Grenoble and also D. Cousins 
for careful reading of the manuscript.


\begin{thebibliography}{99}

\bibitem{houches} For a review on single electron charging efects, 
see "Single Charge Tunneling, Coulomb Blockade Phenomena in 
nanostructures", ed. H. Grabert and M. Devoret, NATO ASI series B: 
Physics Vol {\bf 294} (1992), Plenum Press.

\bibitem{Meso}"Mesoscopic Physics and Electronics", Nanoscience and Technology, 
Ed. Ando et al., Springer Verlag, Berlin Heifdelberg (1998).

\bibitem{bolometers} P. J. Burke, R. J. Schoelkopf and D. E. Prober, J.  
Appl. Phys., Vol. 85, No. 3, (1999); B. Bumble and H. G. LeDuc, IEEE Transactions 
on Applied Superconductivity, Vol. {\bf 7}(2), (1997).

\bibitem{Dolan} G. J. Dolan and J. H. Dunsmuir, Physica B {\bf 152}, 
7-13 (1988).

\bibitem {Beaumont} S. P. Beaumont, P. G. Bower, T. Tamamura and C. D. W. Wilkinson, Appl. Phys. Lett. {\bf 38}, 436 (1981).

\bibitem {Howard} R. E. Howard, E. L. Hu and L. D. Jackel, I.E.E.E. 
Transactions on Electron Devices {\bf 28}(11), 1378 (1981).

\bibitem{Courtois} H. Courtois, P. Gandit and B. Pannetier, Phys. Rev. B 
{\bf 51}, 9360 (1995).

\bibitem{oxygen} M. Strongin, C. Varmazis and A. Joshi, {\it Metallurgy of 
superconducting materials}, in the collection "Treatise on materials and 
technology" Vol.  14, ed. T. Luhman and Dew-Hughes (Academic Press, Inc. 1979), 
pp. 327, 347; See also J. Halbritter, Appl. Phys. A {\bf 43}, 1-28 (1987).

\bibitem{vdz91} H. S. J. van der Zant, H. A. Rijken and J. E. Mooij, 
J. Low Temp. Phys.  {\bf 79}, 289 (1990).

\bibitem{Harada} Y. Harada, D. B. Haviland, P.  
Delsing, C. D. Chen, and T. Cleason, Appl. Phys. Lett. {\bf 65}, 
636 (1994).

\bibitem {Moreau} Wayne M Moreau, Semiconductor 
Lithography (Microdevices), Plenum Press, New York and London (1988) 

\bibitem{Paniez} P. J.  Paniez, J. A.  Guinot, B.  Mortini, C.  
Rosilio, Microelectronic Engineering 41/42, 367-370 (1998) ; See also 
J.  A.  Guinot and P.  J.  Paniez, CNET DTM/MAD/97-35 internal report (1997).

\bibitem{electrosensitivity} In the 
described process, only the high resolution top resist layer is 
electron sensitive. The etching of the bottom layer is achieved by 
either wet or dry etching. However we have noticed a very small 
negative electron sensitivity. No specific step takes this effect 
into account.  

\bibitem{Beenakker} See for example P. W.  Brower and C.  
Beenakker, {\it Chaos and Quantum Transport in Mesoscopic Cosmos}, 
Chaos, Solitons and Fractals, Pergamon/Elsevier, (1997). 

\bibitem{dubos} P.  Dubos et al, to be published.

\end{thebibliography}
\end{document}